\documentclass[journal,letterpaper]{IEEEtran}

\usepackage[cmex10]{amsmath}
\usepackage{amssymb}
\usepackage{cite}
\usepackage{graphicx}

\title{Jensen divergence based on Fisher's information}
\author{Pablo S\'anchez-Moreno, Alejandro Zarzo and Jes\'us S. Dehesa%
\thanks{P. S\'anchez-Moreno is with the Department of Applied Mathematics and the Institute Carlos I for Theoretical and Computational Physics, University of Granada, Granada, Spain}%
\thanks{A. Zarzo is with Department of Applied Mathematics, Polytechnic University of Madrid, Madrid, Spain, and the Institute Carlos I for Theoretical and Computational Physics, University of Granada, Granada, Spain}%
\thanks{J.S. Dehesa is with the Department of Atomic, Molecular and Nuclear Physics and the Institute Carlos I for Theoretical and Computational Physics, University of Granada, Granada, Spain}%
}

\begin{document}

\maketitle

\begin{abstract}

The measure of Jensen-Fisher divergence between probability distributions is introduced and its theoretical grounds set up.
This quantity, in contrast to the remaining Jensen divergences, is
very sensitive to the fluctuations of the probability distributions
because it is controlled by the (local) Fisher information, which is a
gradient functional of the distribution. So, it is appropriate
and informative when studying the similarity of distributions, mainly for those having
oscillatory character.
The new Jensen-Fisher divergence shares with the Jensen-Shannon
divergence the following properties: non-negativity, additivity when
applied to an arbitrary number of probability densities, symmetry
under exchange of these densities, vanishing if and only if all the
densities are equal, and definiteness even when these densities
present non-common zeros. Moreover, the Jensen-Fisher divergence is
shown to be expressed in terms of the relative Fisher information as the
Jensen-Shannon divergence does in terms of the Kullback-Leibler or
relative Shannon entropy.
Finally the Jensen-Shannon and Jensen-Fisher divergences are compared
for the following three large, non-trivial and qualitatively different
families of probability distributions: the sinusoidal, generalized gamma-like and Rakhmanov-Hermite distributions.
\end{abstract}


\section{Introduction}

The study of the measures of similarity between probability densities
is a fundamental topic in probability theory and statistics
\textit{per se} and because of its numerous applications and usefulness in a wide variety of scientific fields, including statistical physics, quantum chemistry, sequence analysis, pattern recognition, diversity, homology, neural networks, computational linguistics, bioinformatics and genomics, atomic and molecular physics and quantum information. The most popular measure of similarity between two probability densities $\rho_1(x)$ and $\rho_2(x)$ is possibly the Jensen-Shannon divergence \cite{lin:itit91,rao_iln87}, which is  defined as
\begin{equation}
JSD[\rho_1,\rho_2]=S\left[\frac{\rho_1+\rho_2}{2}\right]-\frac{S[\rho_1]+S[\rho_2]}{2},
\label{eq_jsd_definition}
\end{equation}
where $S[\rho]$ denotes the Shannon entropy of the density $\rho(x)$, $x\in\Delta \subset  \mathbb{R}$, given by
\[
S[\rho]=-\int_\Delta \rho(x)\ln\rho(x)dx.
\]
According to Eq. (\ref{eq_jsd_definition}), the Jensen-Shannon divergence quantifies the Shannon entropy excess of a couple of distributions with respect to the mixture of their respective entropies. It can also be expressed as 
\[
JSD[\rho_1,\rho_2]=KL\left[\rho_1,\frac{\rho_1+\rho_2}{2}\right]+KL\left[\rho_2,\frac{\rho_1+\rho_2}{2}\right],
\]
indicating that the Jensen-Shannon divergence is a symmetrized and smoothed version of the Kullback-Leibler divergence (KLD in short) or relative Shannon entropy (also called Kullback divergence) defined \cite{kullback_ams51,kullback_68} by
\[
KL[\rho_1,\rho_2]=\int_\Delta
\rho_1(x)\ln\frac{\rho_1(x)}{\rho_2(x)}dx.
\]
The Jensen-Shannon divergence as well as the KLD are non-negative and
vanish if and only if the two densities are equal almost everywhere.
Unlike the KLD,  the Jensen-Shannon divergence has two additional
important characteristics: it is always well defined  (in the sense
that it can be evaluated even when $\rho_1$ is not absolutely
continuous with respect to $\rho_2$), and its square root verifies the triangle inequality so that
the square root of  $JSD[\rho_1, \rho_2]$ is a true metric in the
space of probability distributions \cite{schindelin_itit03}.
Furthermore, it admits the generalization to several probability
distributions \cite{lin:itit91} in the following sense: let be a vector $\boldsymbol\omega  =
(\omega_1, \omega_2,\ldots, \omega_N)$ and a set of $N$ probability
densities $\{\rho_j(x)\}_{j=1}^N$; the Jensen-Shannon divergence among these probability densities is given by 
\begin{multline*}
JSD_{\boldsymbol\omega}[\rho_1,\ldots,\rho_N]=S[\omega_1\rho_1+\cdots+\omega_N\rho_N]\\
-\omega_1S[\rho_1]-\cdots-\omega_N S[\rho_N],
\end{multline*}
where the nonnegative numbers $\omega_i > 0$, for $i=1,\ldots, N$,
such that $\sum_i^N\omega_i=  1$, are weights properly chosen to indicate the relative relevance of each density. This is very useful for certain applications such as in bioinformatics, diversity and atomic physics where there are situations in which it is necessary to measure the overall differences of more than two probability distributions. Notice that for $N = 2$ one has
\[
JSD_{\boldsymbol\omega}[\rho_1,\rho_2]=S[\omega_1\rho_1+\omega_2\rho_2]-\omega_1S[\rho_1]-\omega_2 S[\rho_2],
\]
so that it simplifies to the expression (\ref{eq_jsd_definition}) in
the case $\omega_1=\omega_2=\frac12$.

This divergence has been extensively applied in numerous literary,
scientific and technological areas ranging from information theory
\cite{lin:itit91,topsoe_itit00,tsai_itit05}, statistical and quantum
mechanics \cite{lamberti_07} to bioinformatics and genomics
\cite{romanroldan_prl98,simms_pnas09}, atomic physics
\cite{angulo_pa10,antolin_jcp09,chatzisavvas_jcp05,lopezrosa_pra09}
and quantum information \cite{majtey_epjd05,majtey_ijqi10}. Let us
just mention that it has been used as a tool to study EEG records
\cite{pereyra_pa07}, to segment symbolic sequences
\cite{lamberti_pa03}, to measure the complexity of genomic sequences
\cite{romanroldan_prl98,simms_pnas09}, to analyze literary texts and
musical score \cite{zanette_cs07}, to quantify quantum phenomena such
as entanglement and decoherence \cite{majtey_epjd05,majtey_ijqi10} and to understand the complex organization and shell-filling patterns of the many-electron systems all over the periodic table of chemical elements \cite{angulo_pa10,antolin_jcp09,chatzisavvas_jcp05,lopezrosa_pra09}.

Nevertheless, the Jensen-Shannon divergence is, at times,  weakly
informative or even uninformative, mainly because it depends on a
quantity of global character (the Shannon entropy) in the sense that
it is hardly sensitive to the local fluctuations or irregularities of the probability
densities. So, by definition, this divergence has serious defects to
compare probability densities with highly oscillatory character. This
is often the common situation in many fields, such as e.g. in the
quantum-mechanical description of natural phenomena. To illustrate it,
let us consider the simple case of the motion of a particle-in-a-box
(i.e., in the infinite well  $V(x) = 0$, for $0< x < 1$, and $+\infty$ elsewhere)  \cite{galindo_pascual_90}. The stationary states of the particle are characterized by the sinusoidal probability densities 
\begin{equation}
\rho_n(x)=2\sin^2(\pi n x); x\in (0,1),
\label{eq_infinite_well_wave_functions}
\end{equation}
and $\rho_n(x)=0$ when $x\notin(0,1)$, where $n=1,2,\ldots$ indicates the energetic level and label of the state.
The divergence between the $n$-th quantum state $\rho_n(x)$ and the
ground state $\rho_1(x)$ is studied in Figure \ref{fig:infinite1} by
means of the Jensen-Shannon measure $JSD[\rho_n,\rho_1]$ . We observe
that this divergence tends rapidly to a constant, so that it is not
informative enough about the enormous differences between these two
probability densities.

The case of a particle-in-a-box and other cases pointed out later show
the necessity for defining a new divergence to be able to measure the
similarity between two or more oscillating probability densities in a
much more appropriate quantitative form. This is the purpose of our
work: to introduce the Jensen-Fisher divergence, which depends on an
information-theoretic quantity (the Fisher information
\cite{fisher:pcps25,frieden_04}) with a locality property: it is very
sensitive to fluctuations of the density because it is a gradient
functional of it. This is done in Section \ref{sec_jensen_fisher},
where the definition of the new divergence is given and its main
properties are shown. Then, in Section \ref{sec_comparison} the
Jensen-Shannon and Jensen-Fisher divergences are compared in the
framework of an information theoretic plane for various cases properly
chosen to illustrate the relative advantages and disadvantages of
these two quantities; namely, the sinusoidal, generalized gamma and Rakhmanov-Hermite probability distributions. Finally, some conclusions and open problems are given.

\section{The Jensen-Fisher divergence measure}
\label{sec_jensen_fisher}

In this Section we define a new Jensen divergence between probability distributions based on the Fisher informations of these distributions, and we study its main properties. In doing so, we follow a line of research similar to that of Lin \cite{lin:itit91} to derive the Jensen-Shannon divergence.

Let $X$ be a continuous random variable  with probability density $\rho(x)$, $x\in\Delta\subset\mathbb{R}$. The (translationally invariant) Fisher information of  $\rho(x)$ is given \cite{fisher:pcps25,frieden_04} by 
\begin{equation}
F[\rho]=\int_\Delta  \rho(x) \left[\frac{d}{dx}\ln\rho(x)\right]^2 dx,
\label{eq_fisher_information_definition}
\end{equation}
and  the relative Fisher information between the probability densities $\rho_1(x)$ and $\rho_2(x)$ is defined \cite{hammad:rsa78} by the directed divergence
\[
F_{\rm rel}[\rho_1,\rho_2]=\int_\Delta  \rho_1(x) \left[\frac{d}{dx}\ln\frac{\rho_1(x)}{\rho_2(x)}\right]^2 dx.
\]

It is known that this quantity is non-negative and additive but non symmetric. The relative symmetric measure defined by 
\begin{multline*}
G[\rho_1,\rho_2]=F_{\rm rel}[\rho_1,\rho_2]+F_{\rm
rel}[\rho_2,\rho_1]\\
=\int_\Delta  \left(\rho_1(x)+\rho_2(x)\right)
\left[\frac{d}{dx}\ln\frac{\rho_1(x)}{\rho_2(x)}\right]^2 dx,
\end{multline*}
is called Fisher divergence \cite{hammad:rsa78}, which has been
recently used in some applications to study the complexity and shell
organization of the atomic systems along the Periodic Table
\cite{antolin_jcp09,lopezrosa_pra09}. As in the Shannon case, this
divergence is nonnegative and it vanishes if and only if $\rho_1(x) = \rho_2(x)$ for any $x \in\Delta$, but it is undefined unless that $\rho_1(x)$ and $\rho_2(x)$ be absolutely continuous with respect to each other.

To overcome these problems of the $F_{\rm rel}$ and $G$ divergences, we define a new directed divergence between the probability densities $\rho_1(x)$ and $\rho_2(x)$ as
\[
\overline{F}_{\rm rel}[\rho_1,\rho_2]=\int_\Delta \rho_1(x) \left(\frac{d}{dx}\ln\frac{\rho_1(x)}{\frac{\rho_1(x)+\rho_2(x))}{2}}\right)^2dx.
\]
This  quantity is nonnegative because it
vanishes if and only if $\rho_1(x) = \rho_2(x)$ for any $x\in\Delta$, and it is
well defined even when both densities have non-common zeros. In
addition, it can be expressed in terms of the relative Fisher
information as
\[
\overline{F}_{\rm rel}[\rho_1,\rho_2]=F_{\rm rel} \left[\rho_1,\frac{\rho_1+\rho_2}{2}\right].
\]
However, it is nonsymmetric. To avoid this problem we propose the following symmetrized form 
\begin{equation}
  JFD[\rho_1,\rho_2]=\overline{F}_{\rm
  rel}\left[\rho_1,\rho_2\right]+\overline{F}_{\rm
  rel}\left[\rho_2,\rho_1\right],
\label{eq_jfd_definition}
\end{equation}
as a new measure, which we call \textbf{Jensen-Fisher divergence} between the probability densities $\rho_1(x)$ and $\rho_2(x)$. From Eqs. (\ref{eq_jfd_definition}) and (\ref{eq_fisher_information_definition}), we have that 
\begin{multline*}
  JFD[\rho_1,\rho_2]=F_{\rm rel}
  \left[\rho_1,\frac{\rho_1+\rho_2}{2}\right]+F_{\rm rel}\left[\rho_2,\frac{\rho_1+\rho_2}{2}\right]\\
=\frac12 \left(\int_{-\infty}^\infty
\rho_1(x)\left(\frac{d}{dx}\ln\frac{\rho_1(x)}{\frac{\rho_1(x)+\rho_2(x)}{2}}\right)^2dx\right.\\
+\left.\int_{-\infty}^\infty \rho_2(x)\left(\frac{d}{dx}\ln\frac{\rho_2(x)}{\frac{\rho_1(x)+\rho_2(x)}{2}}\right)^2dx\right)\\
=\frac12\left(\int_{-\infty}^\infty\rho_1(x)\left(\frac{\rho_1'(x)}{\rho_1(x)}-\frac{\rho_1'(x)+\rho_2'(x)}{\rho_1(x)+\rho_2(x)}\right)^2dx\right.\\
+\left.\int_{-\infty}^\infty\rho_2(x)\left(\frac{\rho_2'(x)}{\rho_2(x)}-\frac{\rho_1'(x)+\rho_2'(x)}{\rho_1(x)+\rho_2(x)}\right)^2dx\right)\\
=\frac12\left(
\int_{-\infty}^\infty\left(\frac{(\rho_1'(x))^2}{\rho_1(x)}-2\rho_1'(x)\frac{\rho_1'(x)+\rho_2'(x)}{\rho_1(x)+\rho_2(x)}\right.\right.\\
+\left.\rho_1(x)\frac{(\rho_1'(x)+\rho_2'(x))^2}{(\rho_1(x)+\rho_2(x))^2}\right)dx\\
+
\int_{-\infty}^\infty\left(\frac{(\rho_2'(x))^2}{\rho_2(x)}-2\rho_2'(x)\frac{\rho_1'(x)+\rho_2'(x)}{\rho_1(x)+\rho_2(x)}\right.\\
+\left.\left.\rho_2(x)\frac{(\rho_1'(x)+\rho_2'(x))^2}{(\rho_1(x)+\rho_2(x))^2}\right)dx\right)\\
=
\frac12\int_{-\infty}^\infty \left(\frac{(\rho_1'(x))^2}{\rho_1(x)}+\frac{(\rho_2'(x))^2}{\rho_2(x)}
-2\frac{(\rho_1'(x)+\rho_2'(x))^2}{\rho_1(x)+\rho_2(x)}\right.\\
+\left.\frac{(\rho_1'(x)+\rho_2'(x))^2}{\rho_1(x)+\rho_2(x)}\right)\\
=
\frac12\int_{-\infty}^\infty \left(\frac{(\rho_1'(x))^2}{\rho_1(x)}+\frac{(\rho_2'(x))^2}{\rho_2(x)}
-\frac{(\rho_1'(x)+\rho_2'(x))^2}{\rho_1(x)+\rho_2(x)}\right),
\end{multline*}
so that the Jensen-Fisher divergence can be expressed in terms of the Fisher information as 
\begin{equation}
JFD[\rho_1,\rho_2]=\frac{F[\rho_1]+F[\rho_2]}{2}-F\left[\frac{\rho_1+\rho_2}{2}\right],
\label{eq_jfd_with_fisher}
\end{equation}
which is similar to the expression (\ref{eq_jsd_definition}) of the
Jensen-Shannon divergence in terms of the Shannon entropy, save for a
global minus sign.

It is important to remark that the Jensen-Fisher divergence we have
just introduced, shares the following properties with the Jensen-Shannon divergence. First, it is nonnegative because of Eq. (\ref{eq_jfd_with_fisher}) and the convexity of the Fisher information which leads to 
\[
\frac{F[\rho_1]+F[\rho_2]}{2}\ge F\left[\frac{\rho_1+\rho_2}{2}\right].
\]
Second, it vanishes if and only if the two involved densities are equal almost everywhere in the interval $\Delta$. This comes again from the fact that
\[
\frac{F[\rho_1]+F[\rho_2]}{2}= F\left[\frac{\rho_1+\rho_2}{2}\right] \iff \rho_1=\rho_2,
\]
by keeping in mind the convexity of the Fisher functional; so that,
\[
JFD[\rho_1,\rho_2]=0 \iff \rho_1=\rho_2.
\]
Third, it is symmetric because one can straightforwardly prove that
$JFD[\rho_1, \rho_2]  = JFD[\rho_2, \rho_1]$. Fourth, it is well
defined when $\rho_1(x)$  or $\rho_2(x)$  are not absolutely
continuous with respect to each other (of course, as long as the
Fisher information of each density is well defined) and so, it can be
used to compare probability distributions with no common zeros.

In addition, the Jensen-Fisher and Jensen-Shannon divergences satisfy the following deBruijn-type expression 
\begin{equation}
\left.\frac{d}{d\epsilon}JSD[\rho_1+\sqrt{\epsilon}\rho_G,\rho_1+\sqrt{\epsilon}\rho_G]\right|_{\epsilon=0}=-\frac12 JFD[\rho_1,\rho_2],
\label{eq_de_bruijn_jsd_jfd}
\end{equation}
where $\rho_G$ is a normal distribution with zero mean and variance
equal to one. This can be proved by considering the original
deBruijn's \cite{cover_91} identity between the Shannon entropy and the Fisher information:
\begin{equation}
\left.\frac{d}{d\epsilon}S[\rho+\sqrt{\epsilon}\rho_G]\right|_{\epsilon=0}=\frac12 F[\rho].
\label{eq:debruijn}
\end{equation}
Then,
\begin{multline*}
\left.\frac{d}{d\epsilon}JSD[\rho_1+\sqrt{\epsilon}\rho_G,\rho_2+\sqrt{\epsilon}\rho_G]\right|_{\epsilon=0}\\
=\left.\frac{d}{d\epsilon}S\left[\frac{\rho_1+\sqrt{\epsilon}\rho_G+\rho_2+\sqrt{\epsilon}\rho_G}{2}\right]\right|_{\epsilon=0}\\
-\frac12\left.\frac{d}{d\epsilon}S[\rho_1+\sqrt{\epsilon}\rho_G]\right|_{\epsilon=0}-\frac12\left.\frac{d}{d\epsilon}S[\rho_2+\sqrt{\epsilon}\rho_G]\right|_{\epsilon=0}\\
=\left.\frac{d}{d\epsilon}S\left[\frac{\rho_1+\rho_2}{2}+\sqrt{\epsilon}\rho_G\right]\right|_{\epsilon=0}\\
-\frac12\left.\frac{d}{d\epsilon}S[\rho_1+\sqrt{\epsilon}\rho_G]\right|_{\epsilon=0}-\frac12\left.\frac{d}{d\epsilon}S[\rho_2+\sqrt{\epsilon}\rho_G]\right|_{\epsilon=0}.
\end{multline*}
Taking into account the deBruijn's identity (\ref{eq:debruijn}), we obtain
\begin{multline*}
\left.\frac{d}{d\epsilon}JSD[\rho_1+\sqrt{\epsilon}\rho_G,\rho_2+\sqrt{\epsilon}\rho_G]\right|_{\epsilon=0}\\
=\frac12 F\left[\frac{\rho_1+\rho_2}{2}\right]-\frac14 F[\rho_1]-\frac14 F[\rho_2]=-\frac12 JFD[\rho_1,\rho_2],
\end{multline*}
and the identity (\ref{eq_de_bruijn_jsd_jfd}) is proved.

Furthermore, like the Jensen-Shannon divergence, it admits a
generalization to $N$ densities with different weights
$\boldsymbol\omega=(\omega_1,\omega_2,\ldots,\omega_N)$, where
$\omega_i\ge 0$, $i=0,1,\ldots,N$, and $\sum_{i=1}^N\omega_i=1$,
\begin{multline*}
JFD_{\boldsymbol\omega}[\rho_1,\ldots,\rho_N]=\omega_1F[\rho_1]+\cdots+\omega_N
S[\rho_N]\\
-F[\omega_1\rho_1(x)+\cdots+\omega_N\rho_N(x)].
\end{multline*}

\begin{figure}
  \centering
\includegraphics[width=8cm]{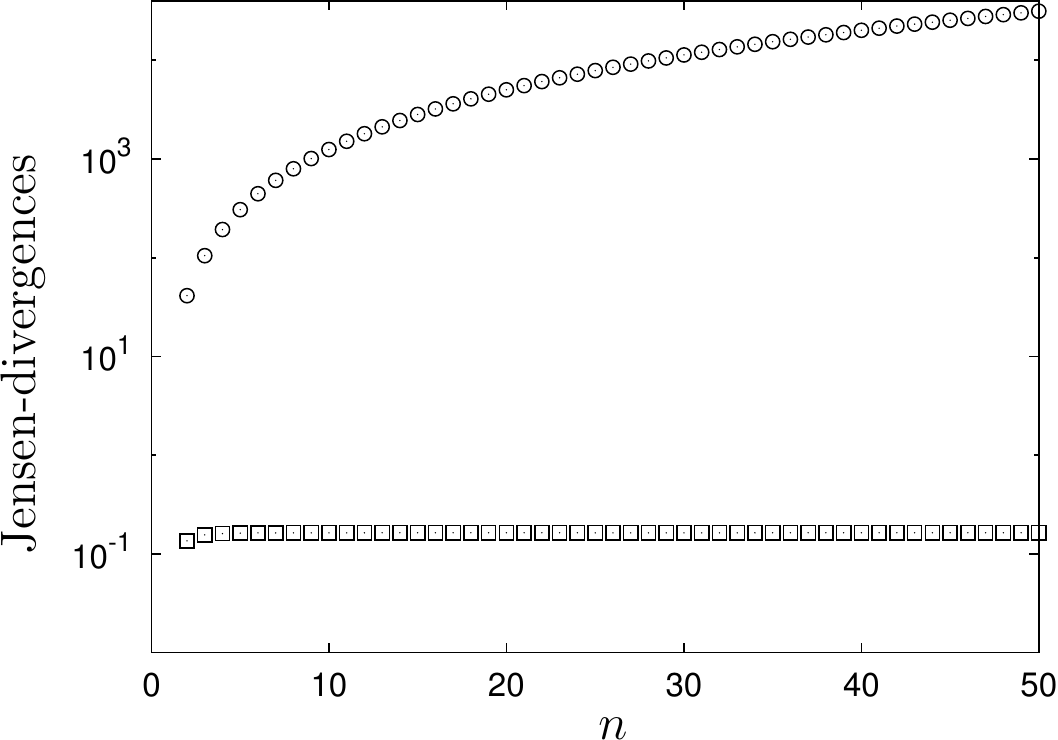}
\caption{Jensen-Shannon $JSD[\rho_n,\rho_1]$ ($\boxdot$) and
Jensen-Fisher $JFD[\rho_n,\rho_1]$ ($\odot$) divergences between the
sinusoidal densities $\rho_n(x)$ and $\rho_1(x)$ (see Eq.
(\ref{eq_infinite_well_wave_functions})) in terms of the quantum number $n$.}
\label{fig:infinite1}
\end{figure}

Finally, let us highlight that the Jensen-Fisher divergence is
informative even in those cases where the Jensen-Shannon is not. This
is illustrated in Figure \ref{fig:infinite1} for the
particle-in-a-box system, whose stationary quantum-mechanical states are described by the
probability densities (\ref{eq_infinite_well_wave_functions}).
Therein, we have depicted the Jensen-Fisher and Jensen-Shannon
divergences between the $n$th-state density $\rho_n(x)$ and the ground
state $\rho_1(x)$, given by $JFD[\rho_n, \rho_1]$ and $JSD[\rho_n,
\rho_1]$ respectively, in terms of $n$ when $n$ is going from 1 to 50.
It turns out that, as $n$ increases, the Jensen-Fisher divergence
increases much more than the
Jensen-Shannon, which remains practically constant. This
clearly indicates that the former divergence is much more informative
than the latter.

\section{Jensen-Shannon and Jensen-Fisher divergences: Mutual comparison}
\label{sec_comparison}

In this Section we compare the Jensen-Fisher and the Jensen-Shannon
divergences in the information-theoretic $JFD-JSD$ plane for the
following three large, qualitatively different families of probability
distributions: the sinusoidal distributions defined by Eq.
(\ref{eq_infinite_well_wave_functions}), the generalized gamma-like
distributions (see Eq. (\ref{eq_gamma_like_def}) below) and the Rakhmanov-Hermite
distributions (see Eq. (\ref{eq_rakhmanov_def}) below).

\subsection{Sinusoidal densities}

These probability densities given by Eq.
(\ref{eq_infinite_well_wave_functions}) have been used to describe
various physical systems, such as e.g. the stationary
quantum-mechanical states of a particle-in-a-box (i.e., in an infinite
potential well) \cite{galindo_pascual_90} as already mentioned.
Indeed, they characterized the ground state $\rho_1(x)$ and the
excited states $\rho_n(x)$, with $n = 2, 3,\ldots$, of this quantum
mechanical system. In Figure
\ref{fig:infinite1}, previously discussed, we have shown that the
excitation  of the particle is described in a much better
information-theoretical way by the $JFD$ than by the $JSD$, since the
former divergence between the probability density of the
$n$th-excited-state and the ground state increases when $n$ (so, when
the energy of the particle) is increasing, while the $JSD$
remains practically constant.

\begin{figure}
  \centering
\includegraphics[width=8cm]{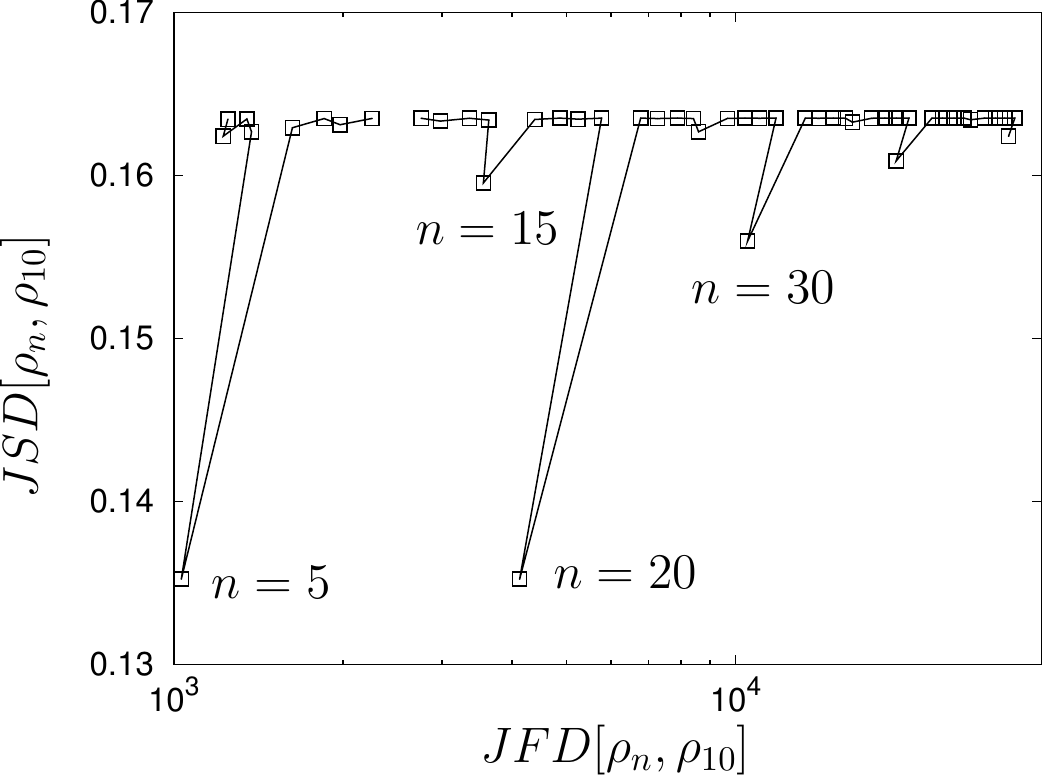}
\caption{$JSD[\rho_n,\rho_{10}]-JFD[\rho_n,\rho_{10}]$ divergence
plane of the sinusoidal densities $\rho_n(x)$ (see Eq.
(\ref{eq_infinite_well_wave_functions})) for $n=1,\ldots,50$.}
\label{fig:infi10}
\end{figure}

In Figure \ref{fig:infi10} we have depicted the Jensen-Shannon
divergence $JSD[\rho_n, \rho_{10}]$ between the excited states with
quantum number $n$ and 10 against the corresponding Jensen-Fisher
divergence $JFD[\rho_n, \rho_{10}]$ for $n = 1,\ldots, 50$. The
resulting values (points) obtained for increasing $n$ are joined by a
line to guide the eye. It is observed that the $JSD$ remains constant except for some points. They correspond to values of $n$ multiple and
submultiple of 10, that is the quantum number of the reference state
$\rho_{10}$. At these points, $\rho_n(x)$ and $\rho_{10}(x)$ share a
number of zeros, so these densities become more similar to each other,
and both $JSD$ and $JFD$ achieve a lower value. Less dramatic
deviations are observed also for values of $n=15,25,35,\ldots$, where
the density $\rho_n(x)$ has some of the zeros of $\rho_{10}(x)$.
From a quantum-mechanical point of view, the particles on those states share some common forbidden regions (or also some common maximum probability regions).
The behaviours of the $JSD$ and $JFD$ measures on this plane shows that although both quantities are sensitive to the overlap of the zeros, the $JFD$ highlights this phenomenon much better (please, be aware of the different scaling in the axes of the figure). Moreover, the $JFD$ presents larger absolute variations and has a much wider range of variation than the $JSD$ along all the pairs of states considered.

\subsection{Generalized gamma-like densities}

In contrast with the previous case (where the densities have several
zeros in a finite interval), here we consider a family of
one-parameter densities having at most one zero and defined in the
whole real line; namely, the gamma-like densities given by
\begin{equation}
\gamma_\beta(x)=\left(\sqrt{2}\,2^\frac{\beta}{2}\Gamma\left(\frac{1+\beta}{2}\right)\right)^{-1}
|x|^\beta\exp\left(-\frac{x^2}{2}\right);\; \beta>1,
\label{eq_gamma_like_def}
\end{equation}
so that for $\beta = 0$, one has a normal distribution:
\[
\gamma(x)\equiv\gamma_0(x)=\frac{1}{\sqrt{2\pi}}\exp\left(-\frac{x^2}{2}\right).
\]
In what follows we assume that $\beta>1$ because the Fisher
information (\ref{eq_fisher_information_definition}) is not defined for $0<\beta\le 1$, having a vertical
asymptote at $\beta=1$.

\begin{figure}
  \centering
\includegraphics[width=8cm]{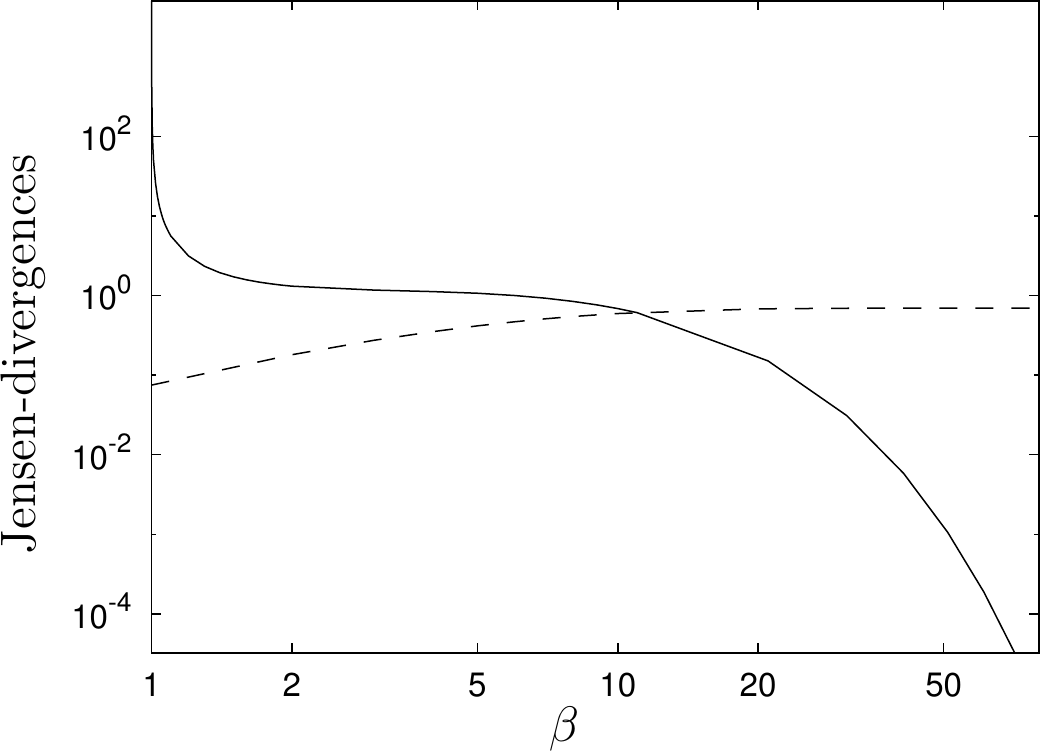}
\caption{Jensen-Shannon $JSD[\gamma_\beta,\gamma]$ (dashed line) and
Jensen-Fisher $JFD[\gamma_\beta,\gamma]$ (solid line) divergences
between the generalized gamma densities $\gamma_\beta(x)$ and
$\gamma(x)$ (see Eq. (\ref{eq_gamma_like_def})) as functions of the multiplicity parameter $\beta$.}
\label{fig:zerobeta}
\end{figure}

We have done two
different analyses. First, in Figure \ref{fig:zerobeta}, the values of
$JFD[\gamma_\beta,\gamma]$ and $JSD[\gamma_\beta,\gamma]$ are given as
a function of $\beta$. It shows that the Jensen-Fisher divergence is much more sensitive to the multiplicity of the zero than the Jensen-Shannon divergence. While the former varies along a range of six orders of magnitude, the latter only varies along one order of magnitude.

\begin{figure}
  \centering
\includegraphics[width=8cm]{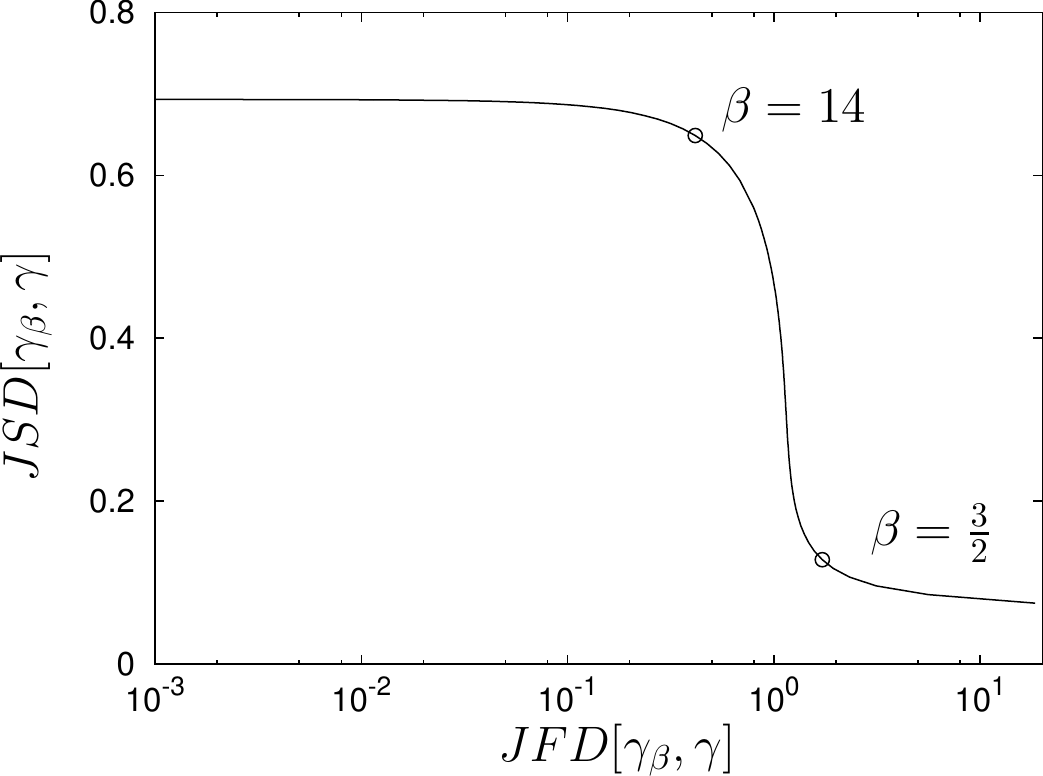}
\caption{$JSD[\gamma_\beta,\gamma]-JFD[\gamma_\beta,\gamma]$
divergence plane of the generalized gamma densities $\gamma_\beta(x)$
and $\gamma(x)$ (see Eq. (\ref{eq_gamma_like_def})) as functions of the multiplicity parameter $\beta$.}
\label{fig:plane_zerobeta}
\end{figure}

Second, Figure \ref{fig:plane_zerobeta} shows the
$JSD[\gamma,\gamma_\beta]-JFD[\gamma,\gamma_\beta]$ plane between the
probability densities $\gamma(x)$ and $\gamma_\beta(x)$ for all values
$\beta$ from 1 to 80. Notice that there are two regimes, one for
$\beta\lesssim \frac32$ and $\beta\gtrsim 14$ where the $JSD$ remains
almost constant and the $JFD$ varies rapidly, and another for $\frac32
\lesssim\beta\lesssim 14$ where the $JFD$ remains almost constant
while the $JSD$ varies. As in the previous example, the range of
variation of the JFD is much wider than that of
$JSD[\gamma,\gamma_\beta]$.

\subsection{Rakhmanov-Hermite densities}

Let us now consider the class of Rakhmanov-Hermite probability densities defined by 
\begin{equation}
\rho^{\rm HO}_n(x)=\frac{1}{2^nn!\sqrt{\pi}}e^{-x^2} H_n^2\left(x\right),
\label{eq_rakhmanov_def}
\end{equation}
where $H_n(x)$ is the orthogonal Hermite polynomial of degree $n$. As
for the quantum infinite well previously discussed, the parameter $n=0,1,2,\ldots$ indicates the energetic level and labels the corresponding state. They have been shown to correspond to the quantum-mechanical probability densities of the ground and excited stationary states of the isotropic harmonic oscillator (HO, in short); see e.g. \cite{nikiforov_88,sanchezmoreno_jcam10}.

\begin{figure}
  \centering
\includegraphics[width=8cm]{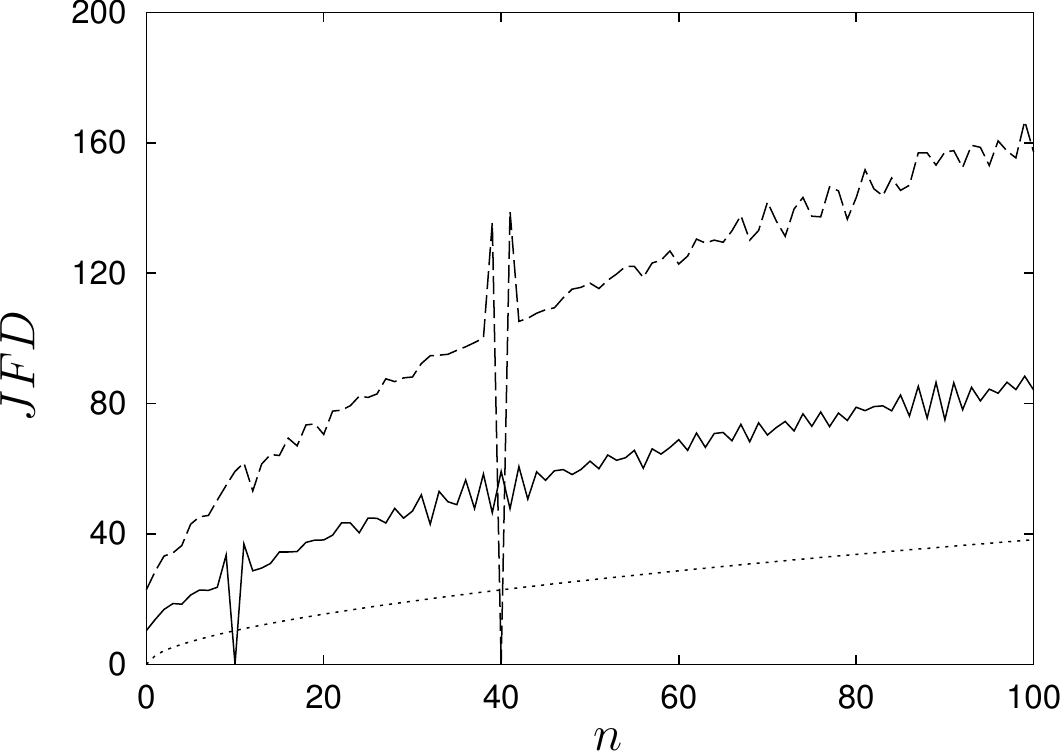}
\caption{Jensen-Fisher divergences $JFD[\rho^{\rm HO}_n,\rho^{\rm
HO}_0]$ (dotted line), $JFD[\rho^{\rm HO}_n,\rho^{\rm HO}_{10}]$
(solid line)  and $JFD[\rho^{\rm HO}_n,\rho^{\rm HO}_{40}]$ (dashed
line) between the $n$th-excited state and the ground state, 10$th$ and
40$th$-excited states of the isotropic harmonic oscillator,
respectively, in terms of the quantum number $n$.}
\label{fig:1dho_jfd}
\end{figure}

\begin{figure}
  \centering
\includegraphics[width=8cm]{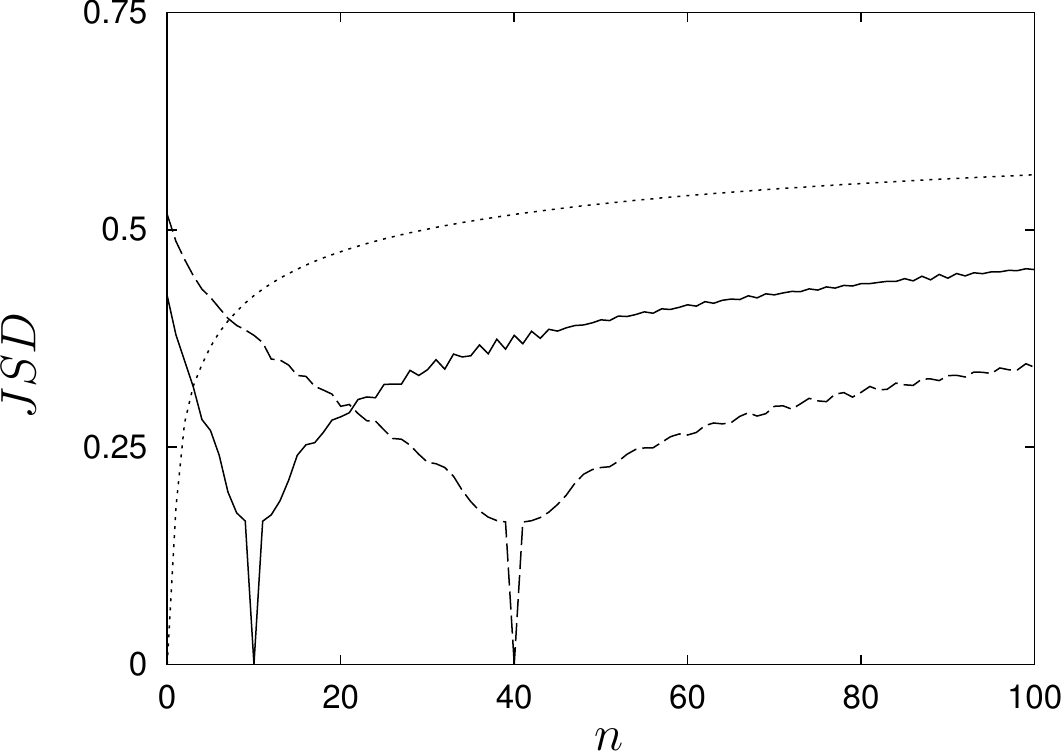}
\caption{Jensen-Shannon divergences $JSD[\rho^{\rm HO}_n,\rho^{\rm
HO}_0]$ (dotted line), $JSD[\rho^{\rm HO}_n,\rho^{\rm HO}_{10}]$
(solid line) and $JSD[\rho^{\rm HO}_n,\rho^{\rm HO}_{40}]$ (dashed
line) between the $n$th-excited state and the ground state, 10$th$ and
40$th$-excited states of the isotropic harmonic oscillator,
respectively, in terms of the quantum number $n$.}
\label{fig:1dho_jsd}
\end{figure}

Here we have done three analyses. Firstly, we depict in Figures
\ref{fig:1dho_jfd} and \ref{fig:1dho_jsd} the Jensen-Fisher and
Jensen-Shannon divergences, respectively, between the $n$th-density
and each of the reference probability densities with $n_r = 0$, $10$
and $40$; this is to say the quantities $JFD[\rho^{\rm HO}_n,\rho^{\rm
HO}_0]$ (dotted line), $JFD[\rho^{\rm HO}_n,\rho^{\rm HO}_{10}]$
(solid line) and $JFD[\rho^{\rm HO}_n,\rho^{\rm HO}_{40}]$ (dashed
line) and the corresponding $JSD$s. We observe from the comparison of
the dotted lines of the two figures that both divergences between the
$n$th-state density $\rho^{\rm HO}_n(x)$ and the ground-state density
$\rho_0(x)$ have a increasing behaviour in terms of the quantum number
$n$ as one should expect. Moreover, from the comparison of the solid
lines of the two figures, we realize an opposite behaviour in the two
divergences between the $n$th-state density $\rho^{\rm HO}_n(x)$ and the
$10$th-state density $\rho^{\rm HO}_{10}(x)$ when the quantum number $n$
(which controls the number of zeros of the density) is increasing;
namely, the $JFD[\rho^{\rm HO}_n,\rho^{\rm HO}_{10}]$ has an
increasing sawtooth behaviour while the $JSD[\rho^{\rm HO}_n,\rho^{\rm
HO}_{10}]$ firstly decreases down to zero when $n$ goes from $0$ to
the reference number $10$, and then increases when $n$ goes from $10$
upwards. A similar trend is observed from the comparison of the dashed
lines of the two figures for the $JFD[\rho^{\rm HO}_n,\rho^{\rm
HO}_{40}]$ and $JSD[\rho^{\rm HO}_n,\rho^{\rm HO}_{40}]$ divergences
but now with respect to the reference number $40$. Clearly, in the three
cases $(n,n_r)=(n,1)$, $(n,10)$ and $(n,40)$ the Jensen-Fisher
divergence has always higher variations than the Jensen-Shannon
divergence, because the $JFD$ has a stronger sensitivity than the $JSD$ to
the increasing oscillatory character of $\rho_n(x)$ when $n$ is
increasing.
In addition
we observe that the Jensen-Fisher divergence presents two maxima
around the reference value (maxima at $9$ and $11$ for $n=10$, and at
$39$ and $41$ for $n=40$) that can be explained taking into account
the relative position of the zeros of the two involved densities. In
those cases each zero of one of the densities is situated between two
zeros of the other density, so none of the zeros of one of the
densities are near the zeros of the other one. The opposite situation
occurs for the local minima that appears in the graphics, where some
zeros of a density are near the zeros of the other one. The Jensen-Shannon
divergence also shows the latter feature but with much less intensity. However, it does not show the local maxima around the reference value.

\begin{figure}
  \centering
\includegraphics[width=8cm]{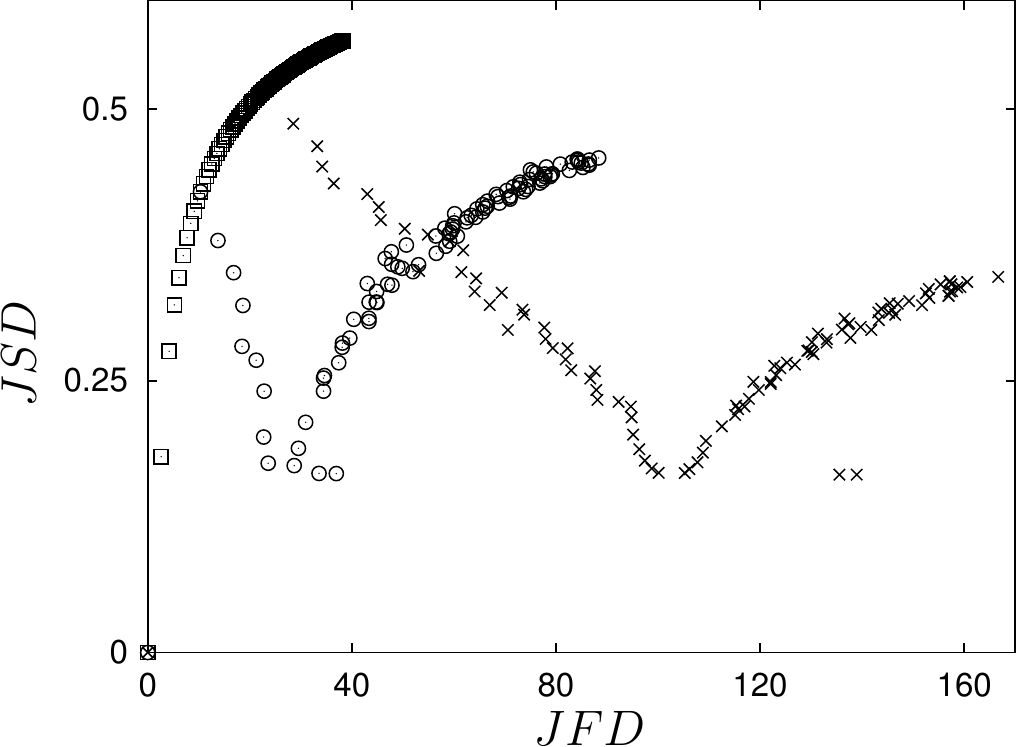}
\caption{$JSD-JFD$ divergence plane of the isotropic harmonic
oscillator for the pairs of stationary states $(n,m)=(n,0)$
($\boxdot$), $(n,10)$ ($\odot$)
and $(n,40)$ ($\times$) when $n$ varies from 0 to 100.}
\label{fig:1dho}
\end{figure}

Our second analysis is shown in Figure \ref{fig:1dho}, where we study the comparison of the $JSD$
and $JFD$ divergences between the pairs of probability densities with
quantum numbers $(n,0)$, $(n, 10)$ and $(n, 40)$ in the frame of the
$JSD-JFD$ divergence plane.  This figure combines the results contained in the two
previous Figures \ref{fig:1dho_jfd}  and \ref{fig:1dho_jsd}, and shows again the overall increasing behaviour of the
Jensen-Fisher divergence and its much higher values, in contrast to
the Jensen-Shannon divergence.
The most important feature that this Figure shows is the
separation of the clouds of points in the direction of increasing $JFD$,
while these clouds are not distinguishable from their $JSD$ values.
Let us mention that the two couples of points
to the right of the vertices of the V-shaped structures, correspond to the local maxima that appear in Figure
\ref{fig:1dho_jfd}.

\begin{figure}
  \centering
\includegraphics[width=8cm]{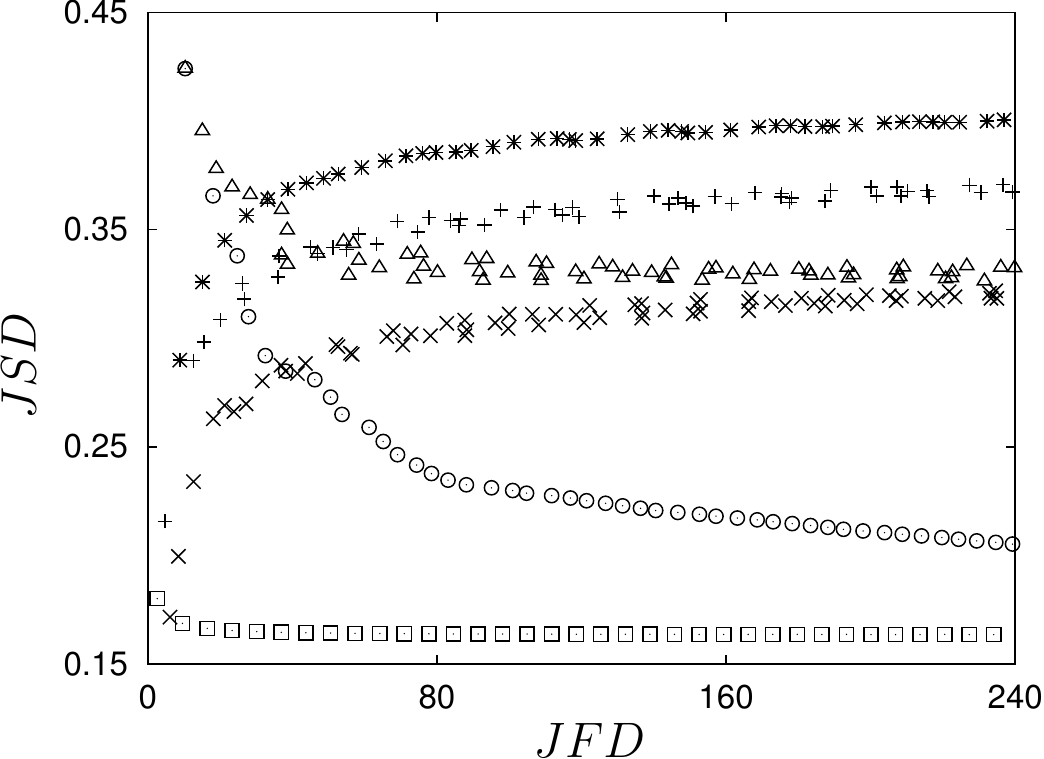}
\caption{$JSD-JFD$ divergence plane of the isotropic harmonic
oscillator for the pairs of stationary states $(n,m)=(n,n+1)$
($\boxdot$),
$(n,n+10)$ ($\odot$), $(n,2n)$ ($\times$), $(n,2n+10)$
($\vartriangle$), $(n,3n)$ ($+$) and $(n,4n)$ ($*$), for several
values of $n$ from $n=0$ up to a value of the $JFD$ of 240.}
\label{fig:1dho_n_n+m}
\end{figure}

Finally, in Figure \ref{fig:1dho_n_n+m} we use again the $JSD-JFD$
plane as a tool to simultaneously show the distance or divergence of
the pairs of probability densities with quantum numbers $(n, n+1)$,
$(n, n+10)$, $(n, 2n)$, $(n, 2n+10)$, $(n, 3n)$ and $(n, 4n)$. Here we
notice that, as $n$ increases, the $JFD$ tends to infinity in all the
cases, but the $JSD$ tends to a constant. We observe that
$JSD[\rho_n^{\rm HO},\rho_{n+1}^{\rm HO}]$ tends to the same value as
$JSD[\rho_n^{\rm HO},\rho_{n+10}^{\rm HO}]$, and  $JSD[\rho_n^{\rm
HO},\rho_{2n}^{\rm HO}]$ tends to the same value as $JSD[\rho_n^{\rm
HO},\rho_{2n+10}^{\rm HO}]$, being those two limiting values different
from each other. Then, we can conclude that this
asymptotic value of the $JSD$ depends on the relative spreading of the
involved densities. When $n$ tends to infinity, the spreading of the
density $\rho_{n}^{\rm HO}(x)$ converges to that of $\rho_{n+1}^{\rm HO}(x)$
or $\rho_{n+10}^{\rm HO}(x)$. However, $\rho_{n}^{\rm HO}(x)$ is less spread
than $\rho_{2n}^{\rm HO}(x)$ or $\rho_{2n+10}^{\rm HO}(x)$. Thus, the JSD
between those densities tends to a different value. This trend is
confirmed by the asymptotic values of 
$JSD[\rho_n^{\rm HO},\rho_{3n}^{\rm HO}]$ and
$JSD[\rho_n^{\rm HO},\rho_{4n}^{\rm HO}]$.

As in previous analyses, Figure
\ref{fig:1dho_n_n+m} shows that the $JFD$ has a
much wider range of variation than the $JSD$, so that it allows us to
discriminate between different values of $n$ in a better way. However,
contrary to what happened in Figure \ref{fig:1dho}, the $JFD$ cannot
distinguish between the different clouds of points of Figure
\ref{fig:1dho_n_n+m}. This is a clear illustration of the
complementarity of both the $JSD$ and $JFD$ when analysing the
similarity of probability distributions. 

\section{Conclusions and open problems}

In this paper the Jensen-Fisher divergence measure is introduced and
its theoretical grounds are shown. In summary, we find that the main
properties (non-negativity, additivity, symmetry, vanishing,
definiteness, deBruijn-like identity) of the Jensen-Shannon divergence
are shared by the new divergence. Moreover, the Jensen-Fisher
divergence is applied to three large families of  representative
probability distributions (sinusoidal, gamma-like and
Rakhmanov-Hermite distributions) and compared with the Jensen-Shannon
divergence.
Our results illustrate that, although both $JSD$ and $JFD$ divergences
are complementary in the sense that they are sensitive to different
aspects of the probability distributions, the latter is more
informative when studying the similarity of oscillating densities.

Finally we should immediately point out three open issues.  First, the
square root of $JSD$ is known to define a metric
\cite{schindelin_itit03,lamberti_pra08}. Does the Jensen-Fisher divergence defines
another distance metric for probability distributions beyond the $JSD$
\cite{schindelin_itit03,topsoe_itit00,tsai_itit05,lamberti_pra08} and
the variational distance \cite{tsai_itit05}?. This is still an open
problem which deserves much attention \textit{per se} and because of
its so many implications in numerous scientific and technological
fields. Second, some generalizations of the $JSD$ have been recently
introduced such as the Jensen-R\'enyi \cite{burbao_itit82} and
Jensen-Tsallis \cite{hamza_jei06,majtey_pa04,lamberti_pa03}
divergences as well as the Jensen divergences of order $\alpha$
\cite{briet_pra09} paying the price of the loss of certain interesting
properties but gaining more flexibility because they have a new degree
of freedom provided by its parameter $q$ or $\alpha$, what is very useful in
numerous applications (see e.g.,
\cite{hamza_03,he_itsp03,chiang_06,wang_09,antolin_jcp10}). Does the
Jensen-Fisher divergence admits any generalization?. The answer is yes
but this avenue is still to be paved. Finally, does there exist a
quantum version of the $JFD$ based on the quantum Fisher information
\cite{luo_lmp00,gibilisco_itit09} similarly to the quantum $JSD$ based
on the von Neuman entropy
\cite{lamberti_pra08,majtey_pra05,lamberti_arxiv09,braunstein_prl94,briet_pra09,roga_prl10}?

\section*{Acknowledgement}

PSM and JSD are very grateful to Junta de Andaluc\'ia for the grants FQM-2445 and FQM-4643,
and the Ministerio de Ciencia e 
Innovaci\'ion for the grant FIS2008-02380. PSM and JSD belong to the research
group FQM-207.

AZ agknowledges partial fiancial support from Ministerio de Educaci\'on
y Ciencia of Spain under grants MTM2006-07186 and MTM2009-14668-C02-02
and from Consejer\'ia de Innovaci\'on, Ciencia y Empresa de la Junta de
Andaluc\'ia, Spain, under grant P09-TEP-5022. Also, AZ has been
partially funded by UPM under some contracts.

This work was finished while on a staying of AZ at Instituo Carlos I
of the Granada University partly funded by this Institute and also by
the Departamento de Matem\'atica Aplicada a la Ingenier\'ia Industrial,
ETSII, UPM.

\bibliographystyle{IEEEtran}
\bibliography{biblio_jensen_fisher}

\end{document}